\documentclass[10pt, conference]{IEEEtran}
\usepackage[utf8]{inputenc}
\usepackage[pdftex,final]{graphicx}
\usepackage{amsmath}
\usepackage{amsthm}
\usepackage{lscape}
\usepackage{latexsym}
\usepackage{amssymb}
\usepackage{bm}
\usepackage{bbm}
\usepackage{cite}
\usepackage{url}
\usepackage{color}
\usepackage[caption=false,font=footnotesize]{subfig}
\usepackage{lmodern}
\usepackage{tikz}
\usepackage{float}
\usepackage[colorinlistoftodos,textwidth=\marginparwidth]{todonotes}
\usetikzlibrary{shapes.geometric, arrows, positioning, fit}

\title{Successive Refinement of Images with \\ Deep Joint Source-Channel Coding}
\author{\IEEEauthorblockN{David Burth Kurka  and  Deniz G\"und\"uz}
\IEEEauthorblockA{Information Processing and Communications Laboratory\\
Department of Electrical and Electronic Engineering\\
Imperial College London, London, UK  \\
{\tt \{d.kurka, d.gunduz\}@imperial.ac.uk}
}
}
\date{}

\begin{document}

\maketitle

\begin{abstract}
    We introduce deep learning based communication methods for successive refinement of images over wireless channels. We present three different strategies for progressive image transmission with deep JSCC, with different complexity-performance trade-offs, all based on convolutional autoencoders. Numerical results show that deep JSCC not only provides graceful degradation with channel signal-to-noise ratio (SNR) and improved performance in low SNR and low bandwidth regimes compared to state-of-the-art digital communication techniques, but can also successfully learn a layered representation, achieving performance close to a single-layer scheme. These results suggest that natural images encoded with deep JSCC over Gaussian channels are almost successively refinable.
\end{abstract}

\section{Introduction}

We consider progressive transmission of images over a point-to-point wireless channel.
In this scenario, an image is transmitted in multiple stages, with each stage improving the quality of the reconstruction. Typically, we expect the first layer to be low-quality, but enough to convey the main elements of the content being transmitted. Following layers are then used to enhance the image quality, by adding more details and components to it~\cite{Sloan1979}.
Progressive transmission can be applied to scenarios in which communication is either expensive or urgent.
For example, in surveillance applications it may be beneficial to quickly send a low-resolution image to detect a potential threat as soon as possible, while a higher resolution description can be later received for further evaluation or archival purposes.
It is also possible that the higher layers can be received by only a subset of the receivers. This may be the case in wireless multicasting of the same image to devices with different resolutions. Progressive transmission would allow low-resolution devices to receive and decode only a limited portion of the channel resources, saving energy, while high-resolution receivers can recover a better quality reconstruction by receiving additional symbols.

Information theoretically this problem corresponds to hierarchical joint source-channel coding (JSCC), studied in \cite{Steinberg2004}, where the optimality of separation is proven; that is, it is optimal to compress the image into multiple layers using successive refinement source coding~\cite{Equitz1991}, where the rate of each layer is dictated by the capacity of the channel it is transmitted over. In general, successive refinement introduces losses compared to single-layer compression at the highest possible resolution; that is, the adaptation to channel bandwidth comes at a price, although some ideal source distributions are known to be \emph{successively refinable} under certain performance measures, which means that they can be progressively compressed at no rate loss, for example, Gaussian sources over Gaussian channels~\cite{Equitz1991}.
On the other hand, it is known that in practical scenarios JSCC can provide gains compared to separate source and channel code design.

Here, following our previous work~\cite{Bourtsoulatze2018a}, we use deep learning (DL) methods, in particular, the autoencoder architecture \cite{GoodfellowDL2016}, for the design of an end-to-end progressive image transmission system.
In~\cite{Bourtsoulatze2018a}, we introduced a novel end-to-end DL-based JSCC scheme for image transmission over wireless channels, called the \emph{deep JSCC}, where encoding and decoding functions are parameterized by convolutional neural networks (CNNs) and the communication channel is incorporated into the neural network (NN) architecture as a non-trainable layer. This method achieves remarkable performance in low signal-to-noise ratio (SNR) and limited channel bandwidth, also showing resilience to mismatch between training and test channel conditions and channel variations similarly to analog communications.

DL-based methods are receiving significant attention for the design of novel and efficient coding and modulation techniques. In particular, the similarities between the autoencoder architecture and the digital communication systems have motivated many studies including decoder design for existing channel codes \cite{Kim:ICLR:18, Nachmani:JSTSP:18}, blind channel equalization \cite{vae:bce}, learning codes for SISO \cite{deep:PHY} and MIMO \cite{deep:MIMO} systems, OFDM systems \cite{JuangWCL2018,Felix:SPAWC:18}, JSCC of text messages \cite{FarsadICASSP2018}, and JSCC for analog storage \cite{ZarconeDCC2018}.
Similar methods have also recently shown notable results in image compression \cite{TodericiICLR2016,BalleICLR2018}%

We propose three alternative architectures for progressive deep JSCC with different complexities. The results are remarkable in the sense that progressive transmission in multiple layers introduces a modest performance loss (in terms of average peak signal-to-noise ratio - PSNR) compared to single-layer transmission; that is, deep JSCC allows adding new layers with almost no penalties on the performance of the existing layers. This result suggests that natural images transmitted with deep JSCC are successively refinable over Gaussian channels under the PSNR performance measure. This also suggests that deep JSCC not only provides natural adaptation to the channel quality~\cite{Bourtsoulatze2018a}, but also to the bandwidth. It is shown in \cite{Bourtsoulatze2018a} that deep JSCC has better or comparable performance to separate source and channel coding (JPEG2000 followed by high performance channel codes) in single-layer transmission. Here we show that the advantages of deep JSCC extend to progressive transmissions as well.

\section{Background and Problem Formulation}

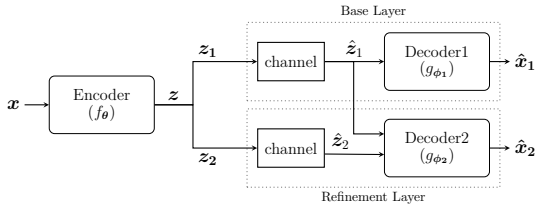
\begin{figure}[t]
	\begin{center}
        \resizebox {0.85\linewidth} {!}%
        {%

 \tikzstyle{txt} = [text centered]
 \tikzstyle{box} = [rectangle, rounded corners, minimum width=2.5cm, minimum height=1.5cm, text width=2.5cm, text centered, draw=black]
 \tikzstyle{bbox} = [rectangle, thick, minimum width=1.5cm, minimum height=1cm, text width=1.5cm, text centered, draw=black]
 \tikzstyle{arrow} = [thick,->,>=stealth]
 \tikzstyle{fitted} = [draw=gray, thick, dotted, inner sep=0.75em]

\begin{tikzpicture}[node distance=2.3cm]

\node (x) [txt, font=\fontsize{14}{0}\selectfont] {$\bm{x}$};
\node (encoder) [box, right of=x, font=\fontsize{12}{12}\selectfont] {Encoder \\ ($f_{\bm{\theta}}$)};
\node (z) [txt, right of=encoder, xshift=0.5cm, font=\fontsize{14}{0}\selectfont] { };
\node (channel1) [bbox, above right of=z,  font=\fontsize{12}{12}\selectfont, xshift=0.5cm, yshift=-0.5cm] {channel};
\node (decoder1) [box, right of=channel1, xshift=1.5cm,  font=\fontsize{12}{12}\selectfont] {Decoder1\\($g_{\bm{\phi_1}}$)};
\node (xhat1) [txt, right of=decoder1,font=\fontsize{14}{0}\selectfont] {$\bm{\hat{x}_1}$};
\node (baselayer) [fitted, fit=(channel1) (decoder1)] {};
\node at (baselayer.north) [above, inner sep=1mm] {Base Layer};

\node (channel2) [bbox, below right of=z,  font=\fontsize{12}{12}\selectfont, xshift=0.5cm, yshift=0.5cm] {channel};
\node (decoder2) [box, right of=channel2, xshift=1.5cm,  font=\fontsize{12}{12}\selectfont] {Decoder2\\($g_{\bm{\phi_2}}$)};
\node (xhat2) [txt, right of=decoder2,font=\fontsize{14}{0}\selectfont] {$\bm{\hat{x}_2}$};
\node (reflayer) [fitted, fit=(channel2) (decoder2)] {};
\node at (reflayer.south) [below, inner sep=1mm] {Refinement Layer};

\draw [arrow] (x) -- (encoder);
\draw [arrow] (encoder.east)
-- node[above,font=\fontsize{14}{0}\selectfont] {$\bm{z}$} ++(1, 0) |- node[above right,font=\fontsize{14}{0}\selectfont] {$\bm{z_1}$}  (channel1.west);
\draw [arrow] (channel1) -- node[above,font=\fontsize{14}{0}\selectfont] {$\hat{\bm{z}}_1$} (decoder1);
\draw [arrow] (decoder1) -- (xhat1);
\draw [arrow] (encoder.east)
-- ++(1, 0) |- node[below right,font=\fontsize{14}{0}\selectfont] {$\bm{z_2}$}  (channel2.west);
\draw [arrow] (channel2.350) -- node[above left,font=\fontsize{14}{0}\selectfont] {$\hat{\bm{z}}_2$} (decoder2.187);
\draw [arrow] (channel1.east) -- ++(0.75, 0) |- (decoder2.165);
\draw [arrow] (decoder2) -- (xhat2);

\end{tikzpicture}

}%
	\end{center}
\vspace{-0.4cm}
\caption{Progressive wireless image transmission with two layers.}
\label{fig:modeldirectpass}
\vspace{-0.3cm}
\end{figure}

We consider progressive wireless transmission of images, where the input image $\bm x\in  \mathbb{R}^n$ is transmitted in $L$ layers. Let ${\bm z_i, \hat{\bm z}_i} \in \mathbb{C}^{k_i}$ denote the complex channel input and output vectors for the $i$th layer, $i \in [L]  \triangleq [1, \dots, L]$. The receiver outputs a different image reconstruction $\hat{\bm x}_i$ after receiving the $i$th layer (using the first $i$ layers). Equivalently, we can consider $L$ virtual receivers corresponding to each layer. See Figure~\ref{fig:modeldirectpass} for an illustration of the system model for $L=2$.
We will call the image dimension $n$ as the \textit{source bandwidth}, and the channel dimension $k_i$ as the \textit{bandwidth of channel $i$}. We will refer to the ratio $k_i/n$ as \textit{bandwidth compression ratio} for the $i$th layer.
An average power constraint is imposed on the transmitted signal at each layer ${\bm z}_i$, $\frac{1}{k_i} \mathbb{E}[{\bm z_i}^*{\bm z_i}] \leq P$.

The reconstruction after receiving the first $i$ layers is denoted by $\hat{\bm x}_i \in \mathbb{R}^n$. Its performance is evaluated by the peak signal to noise ratio ($\mathrm{PSNR}_i$), which is the inverse of the mean square error (MSE), defined as:

\begin{equation*}
\mathrm{MSE}_i = \frac{1}{n} ||\bm x-\hat{\bm x}_i ||^2;~
\mathrm{PSNR}_i = 10\log_{10}\frac{\mathrm{MAX}^2}{\mathrm{MSE}_i},
\end{equation*}
where $\mathrm{MAX}$ is the maximum value a pixel can take, which is $255$ in our case.
Two channel models, the additive white Gaussian noise (AWGN) channel and the slow Rayleigh fading channel, are considered in this work.

\begin{figure}[t]
    \centering
    \includegraphics[width=\linewidth]{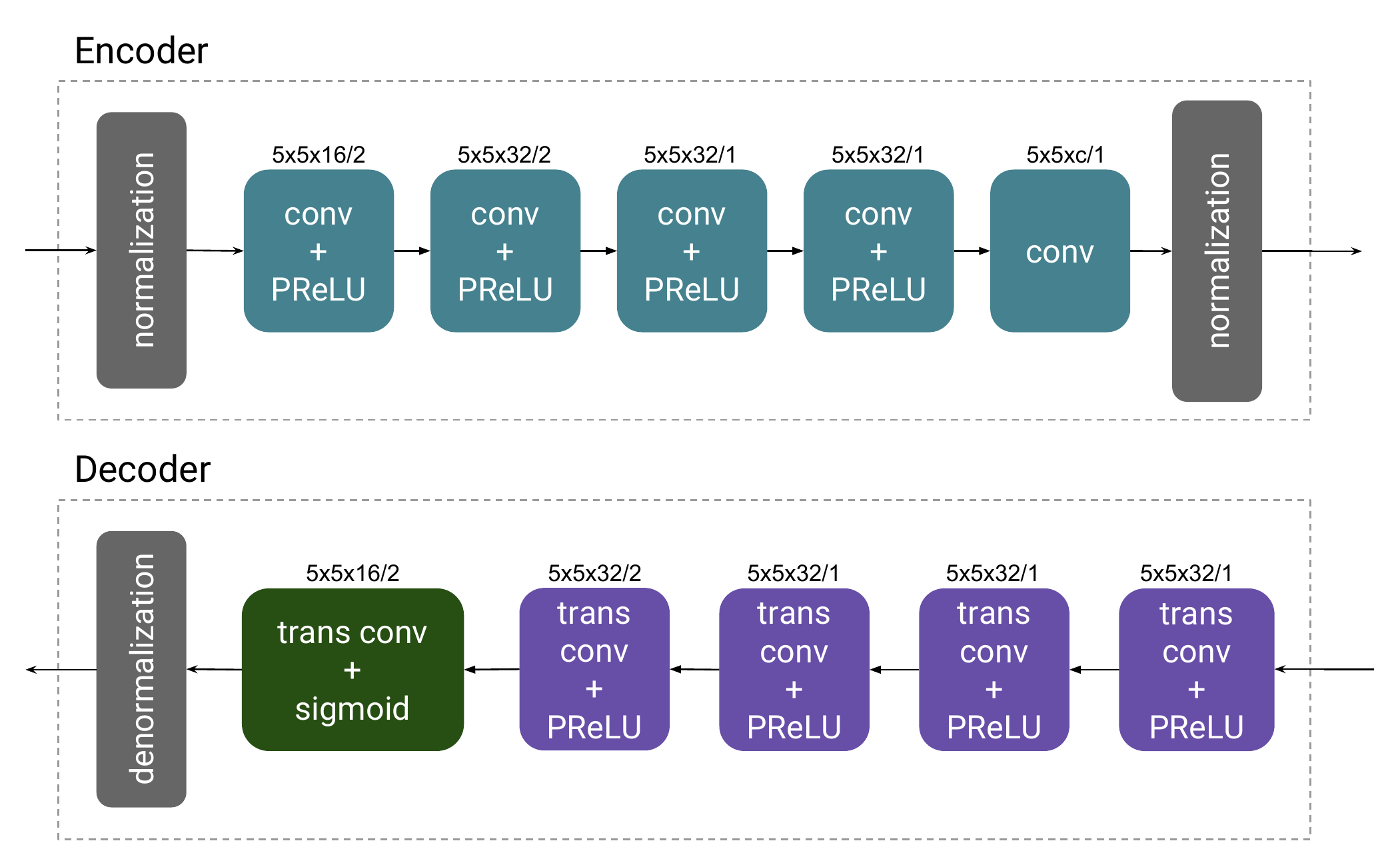}
\vspace{-0.6cm}
    \caption{The encoder and decoder components used in this paper, introduced in \cite{Bourtsoulatze2018a}. The notation $k\times k \times d / s$ refers to kernel size $k$, depth $d$ and stride $s$. $c$ defines the encoder's compression rate.}
    \label{fig:architecture}
\vspace{-0.3cm}
\end{figure}

All the proposed schemes share the same CNN architecture for the encoder and decoder components, shown in Figure~\ref{fig:architecture}.
Once the model is created, the encoder(s) and decoder(s) are trained jointly as unsupervised learning, while the channel is incorporated to the model as a non-trainable layer, producing random values at every realization. We implement our models in TensorFlow and use the Cifar10 dataset in all the experiments. Previous work~\cite{Bourtsoulatze2018a} demonstrated the efficiency of this architecture for JSCC, creating encoders capable of directly mapping pixels to channel inputs, and decoders that retrieve the underlying image directly from the noisy channel outputs, achieving better or comparable performance with the state-of-the-art digital separation-based transmission schemes.
In the next sections, we will present different strategies for progressive JSSC of images, and present numerical results.

\section{Multiple Decoders}
\label{sec:muldec}

\begin{figure*}[ht!]
	\begin{center}
 		\subfloat[]{\includegraphics[height=165pt]{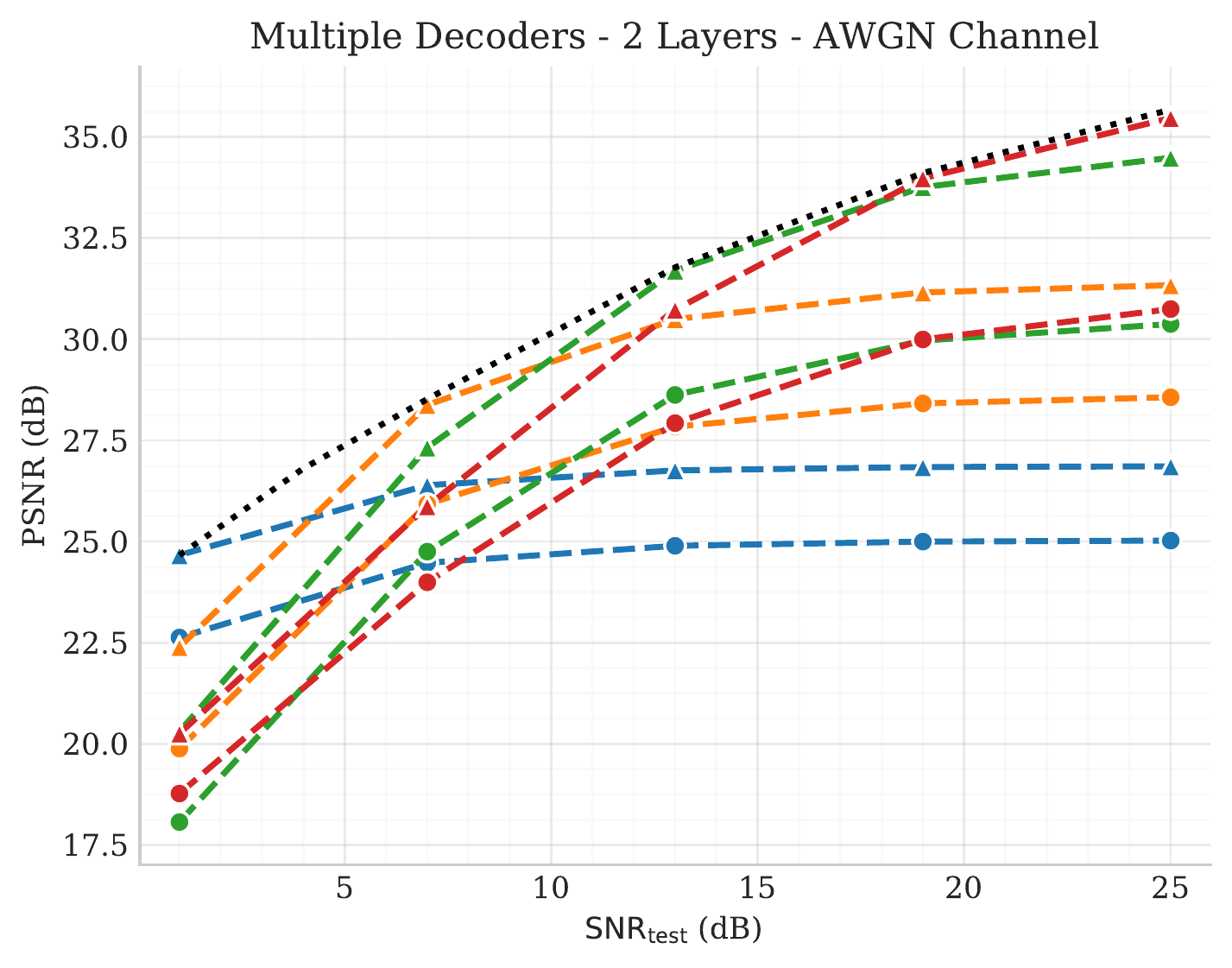} \label{fig:dpass-awgn-05}}
		\subfloat[]{\includegraphics[height= 165pt]{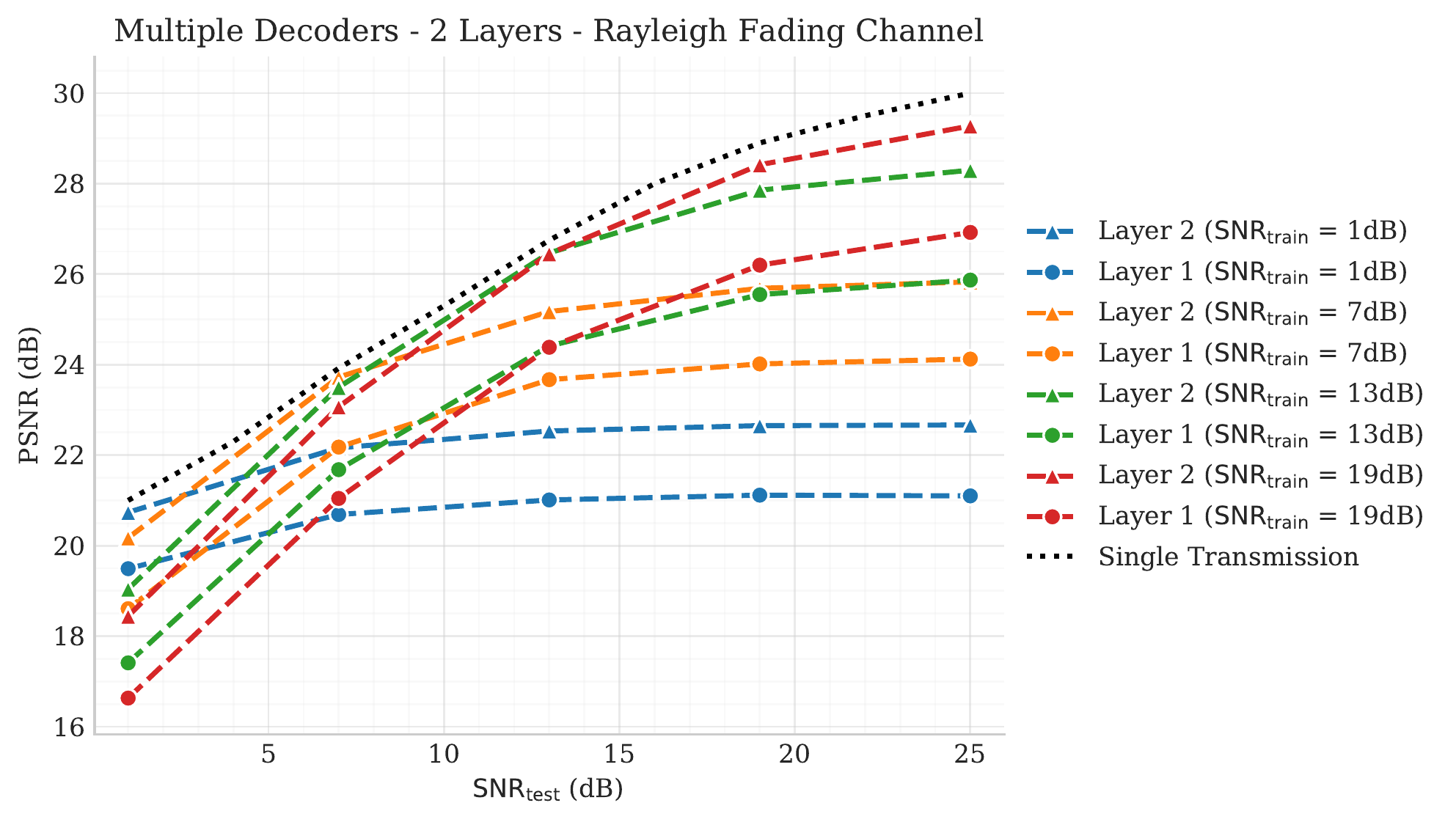}\label{fig:dpass-fading}}
		\\
		\vspace{-0.2cm}
		 \subfloat[]{\includegraphics[height=165pt]{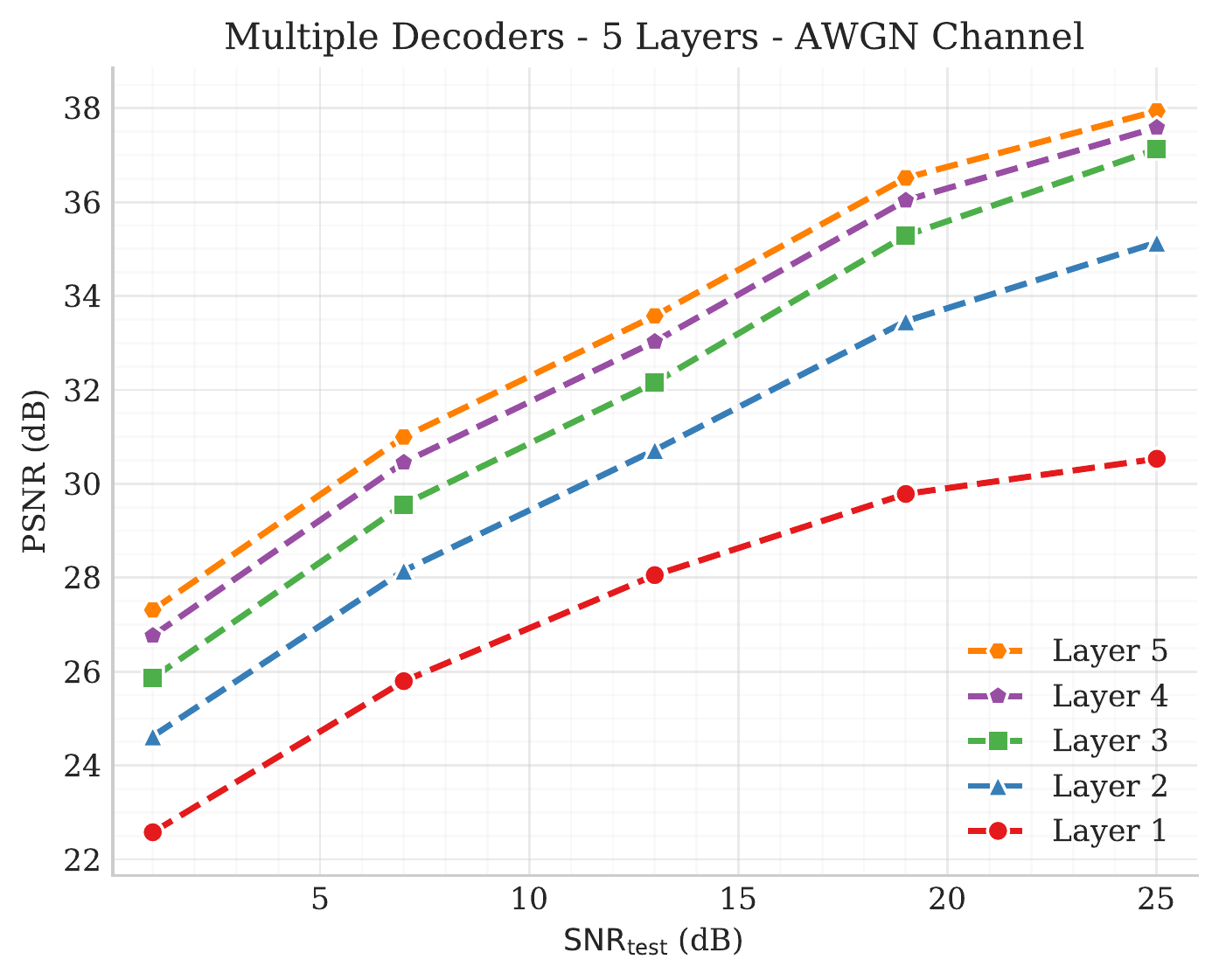}\label{fig:dpass-mult}}
		 \hspace{5pt}
		 \subfloat[]{\includegraphics[height=165pt]{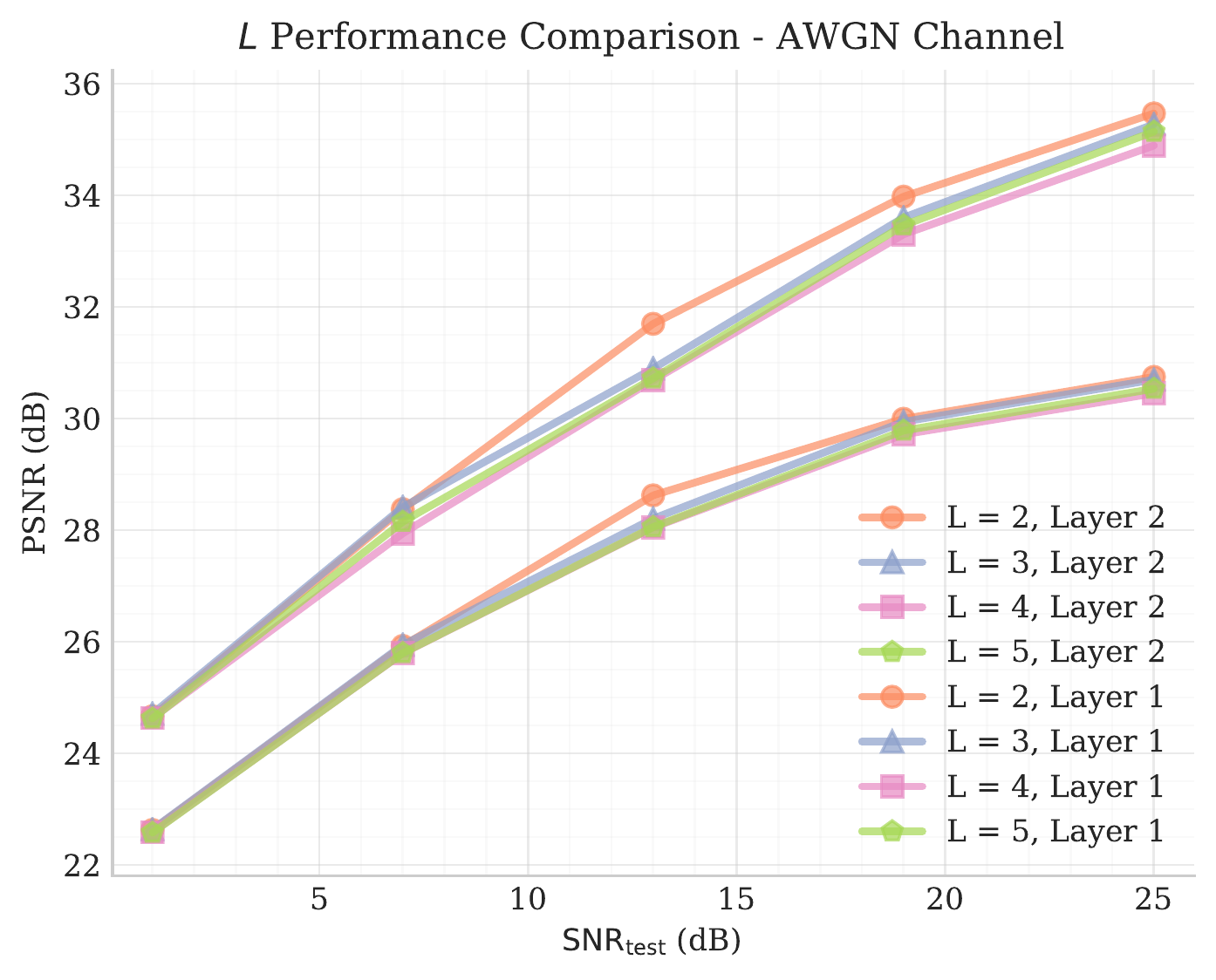}\label{fig:dpass-laycomp}\hspace{50pt}}
	\end{center}
	\vspace{-0.3cm}
		\caption{Performance of multiple decoders scheme on CIFAR-10 test images for (a) AWGN channel; (b) fading channel; (c) AWGN channel, five decoders; (d) two first layers' performance of models trained with different values of $L$. In all images, $k_i/n = 1/12,~ \forall i \in \{1 \dots 5\}$}
		\vspace{-0.3cm}
\end{figure*}

In the first model we consider a single encoder NN generating at once the complete channel input vector ${\bm z} = [{\bm z}_1 \cdots {\bm z}_L]$, which is transmitted in $L$ stages. Then $L$ independent decoder NNs are considered, where the $i$th decoder uses $\sum_{j=1}^i k_j$ channel output symbols to output a distinct reconstruction of the input image.

The system is modelled as an autoencoder with one encoder and $L$ decoder NNs, with the loss defined as:

\begin{equation}
\mathcal{L} = \frac{1}{L}\frac{1}{N}\sum_{i=1}^L\sum_{j=1}^N d(\bm x_j, \hat{\bm x}_{j,i}),
\label{eq:lossdirect}
\end{equation}
where $d(\bm x, \hat{\bm x})$ is the MSE distortion and $N$ the number of training samples.

\subsection{Two-layer model}

We first focus on the ($L=2$) layers scenario which requires the training of only one encoder and two decoders. The second decoder receives the output of both transmissions; and hence, should achieve a better performance.

\subsubsection{AWGN Channel}

We consider a model with $k_1/n = k_2/n = 1/12$, transmitting over an AWGN channel. We experiment different channel qualities (specified by the channel SNR) and train different models for different target SNR. Experimental results are shown in Figure~\ref{fig:dpass-awgn-05}, where each colour represents the performance of a model trained for a specific SNR, with one curve corresponding to the performance of the decoder receiving only the base layer, while the other to the decoder receiving both layers.
Although a model is optimized for a specific SNR, our results show, for each trained model, evaluations in a range of test SNRs (1-25dB).
We see that in all cases, average $\mathrm{PSNR}_2$ is consistently higher than $\mathrm{PSNR}_1$ by 2 to 3 dB, showing that the successive refinement has been achieved.

As a baseline, we compare our results to a single layer transmission scheme, with the same channel bandwidth as the sum of the individual layers, that is $k=k_1+k_2$.
We see that the progressive JSCC scheme can approach the same performance as the single layer, showing that there
are no significant losses in the transmission efficiency when the model is adapted for successive refinement.

The evaluation in multiple SNRs, including those lower than the trained SNR, shows that the scheme is robust against channel deterioration, not suffering from the \emph{cliff effect}, and instead presenting graceful degradation. This analog property of the model was already observed in the single layer case in \cite{Bourtsoulatze2018a}. This behaviour is valid for all other results presented in this paper; however, due to a space limitation, those results will not be explicitly shown.

\subsubsection{Fading channel}

We then consider the same model on a slow Rayleigh fading channel.
Figure~\ref{fig:dpass-fading} shows results for an architecture similar to the one used in Figure~\ref{fig:dpass-awgn-05}. We see that, although the PSNRs are lower than those in the AWGN case due to channel variations, the overall properties of graceful degradation and analog behaviour are still present. Besides, the performance of the deep-JSCC scheme is significantly superior to the state-of-art separation based schemes in the case of fading channels, despite the lack of explicit pilots or channel estimation~\cite{Bourtsoulatze2018a}.

Although all the models exhibit similar behaviour over fading channels, we will limit our attention to the AWGN channel in the rest of the paper due to the space limitation.

\subsection{Multiple layers}

Next, we extend the model to more layers. Figure~\ref{fig:dpass-mult} shows the results for $L=5$ layers, each with bandwidth compression equal to $1/12$. For each test SNR, only the highest PSNR obtained is plotted (i.e., the convex hull of the previous plots).
The results show that the addition of new layers increases the overall quality of the transmitted image at every step; although the amount of improvement is diminishing, as the model is able to transmit the main image features with the lower layers, leaving only marginal contributions to the additional layers.

We also notice that the introduction of additional layers in the training model has very low impact on the performance of the first layers, compared to models with smaller values of $L$. This can be seen in Figure~\ref{fig:dpass-laycomp}, which compares the performance of the first and second layers for models trained with $L \in \{2,3,4,5\}$, showing that the loss of adding new layers is negligible.
This is rather surprising, given that the code of the first layer is shared by all the layers and is optimized for all layers, as in Eq.~\ref{eq:lossdirect}. The results, therefore, suggest that there is performance independence between layers, justifying the use of as many layers as desired, as long as there are available resources.

\section{Residual Transmission}

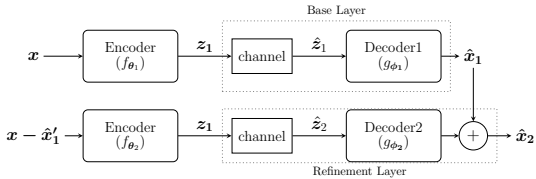
\begin{figure}[h]
	\begin{center}
        \resizebox {0.85\linewidth} {!}%
        {%

 \tikzstyle{txt} = [text centered]
 \tikzstyle{box} = [rectangle, rounded corners, minimum width=2.5cm, minimum height=1.5cm, text width=2.5cm, text centered, draw=black]
 \tikzstyle{bbox} = [rectangle, thick, minimum width=1.5cm, minimum height=1cm, text width=1.5cm, text centered, draw=black]
 \tikzstyle{arrow} = [thick,->,>=stealth]
 \tikzstyle{fitted} = [draw=gray, thick, dotted, inner sep=0.75em]
 \tikzstyle{sum} = [draw, circle]

\begin{tikzpicture}[node distance=2.3cm]

\node (x1) [txt, font=\fontsize{14}{0}\selectfont] {$\bm{x}$};
\node (encoder1) [box, right of=x1, xshift=0.5cm, font=\fontsize{12}{12}\selectfont] {Encoder \\ ($f_{\bm{\theta}_1}$)};
\node (channel1) [bbox, right of=encoder1, xshift=1cm, font=\fontsize{12}{12}\selectfont, xshift=0.5cm] {channel};
\node (decoder1) [box, right of=channel1, xshift=1.5cm,  font=\fontsize{12}{12}\selectfont] {Decoder1\\($g_{\bm{\phi_1}}$)};
\node (xhat1) [txt, right of=decoder1,font=\fontsize{14}{0}\selectfont] {$\bm{\hat{x}_1}$};
\node (baselayer) [fitted, fit=(channel1) (decoder1)] {};
\node at (baselayer.north) [above, inner sep=1mm] {Base Layer};

\node (x2) [txt, below of=x1, font=\fontsize{14}{0}\selectfont] {$\bm{x-\hat{x}'_1}$};
\node (encoder2) [box, right of=x2, xshift=0.5cm, font=\fontsize{12}{12}\selectfont] {Encoder \\ ($f_{\bm{\theta}_2}$)};
\node (channel2) [bbox, right of=encoder2, xshift=1cm, font=\fontsize{12}{12}\selectfont, xshift=0.5cm] {channel};
\node (decoder2) [box, right of=channel2, xshift=1.5cm, font=\fontsize{12}{12}\selectfont] {Decoder2\\($g_{\bm{\phi_2}}$)};
\node (adder) [sum, right of=decoder2] {\Large$+$};
\node (xhat2) [txt, right of=adder, xshift=-0.8cm, font=\fontsize{14}{0}\selectfont] {$\bm{\hat{x}_2}$};
\node (reflayer) [fitted, fit=(channel2) (adder)] {};
\node at (reflayer.south) [below, inner sep=1mm] {Refinement Layer};

\draw [arrow] (x1) -- (encoder1);
\draw [arrow] (encoder1) -- node[above,font=\fontsize{14}{0}\selectfont] {$\bm{z_1}$} (channel1);
\draw [arrow] (channel1) -- node[above,font=\fontsize{14}{0}\selectfont] {$\hat{\bm{z}}_1$} (decoder1);
\draw [arrow] (decoder1) -- (xhat1);
\draw [arrow] (x2) -- (encoder2);
\draw [arrow] (encoder2) -- node[above,font=\fontsize{14}{0}\selectfont] {$\bm{z_1}$} (channel2);
\draw [arrow] (channel2) -- node[above,font=\fontsize{14}{0}\selectfont] {$\hat{\bm{z}}_2$} (decoder2);
\draw [arrow] (decoder2) -- (adder);
\draw [arrow] (xhat1) -- (adder);
\draw [arrow] (adder) -- (xhat2);

\end{tikzpicture}

}%
	\end{center}
\vspace{-0.3cm}
\caption{Residual transmission scheme with two layers.}
\label{fig:modelresnet}
\end{figure}

We proceed our investigation by proposing an alternative model scheme. As seen in Figure~\ref{fig:modelresnet}, each transmission is performed by an independent encoder/decoder pair acting in sequence. The first pair (the base layer) is designed to transmit the original image $\bm x$, retrieving $\hat{\bm x}_1$. Then, each subsequent layer computes an estimate of the image reconstructed at the receiver side using all the previous layers, so it can transmit only a residual image:

$${\bm x}_i^{\textit{res}} = {\bm x} - \sum_{j=1}^{i-1} {\hat{\bm x}'_j}~.$$

We assume that the estimated output $\hat{\bm x}_j'$ is equal to the actual receiver output during the training phase (i.e., ${\hat{\bm x}'_j} = {\hat{\bm x}_j}$); however, during evaluation, we consider the receiver is deployed and inaccessible, so the estimation is obtained by averaging independent realizations of the channel and decoder models at the transmitter side (i.e., ${\hat{\bm x}'_l} = \frac{1}{m}\sum_{i=1}^m {\hat{\bm x}}$, where $m$ is the number of independent channel realizations used to estimate receiver's output). %

The main advantage of this scheme is the fact that each encoder/decoder pair can be optimized separately, given the result of the previous layers. Although this is more computationally demanding, it allows design flexibility, as new layers can be added as they are required, without the need of changes on the previously trained parts.
This could be used, for example, in an expanding system that add refinement layers as resources become available, or in a distributed communication settings, in which transmitters located at different regions refine a message, given a receiver's feedback.

\begin{figure}[t]
	\begin{center}
 		\includegraphics[width=0.45\textwidth]{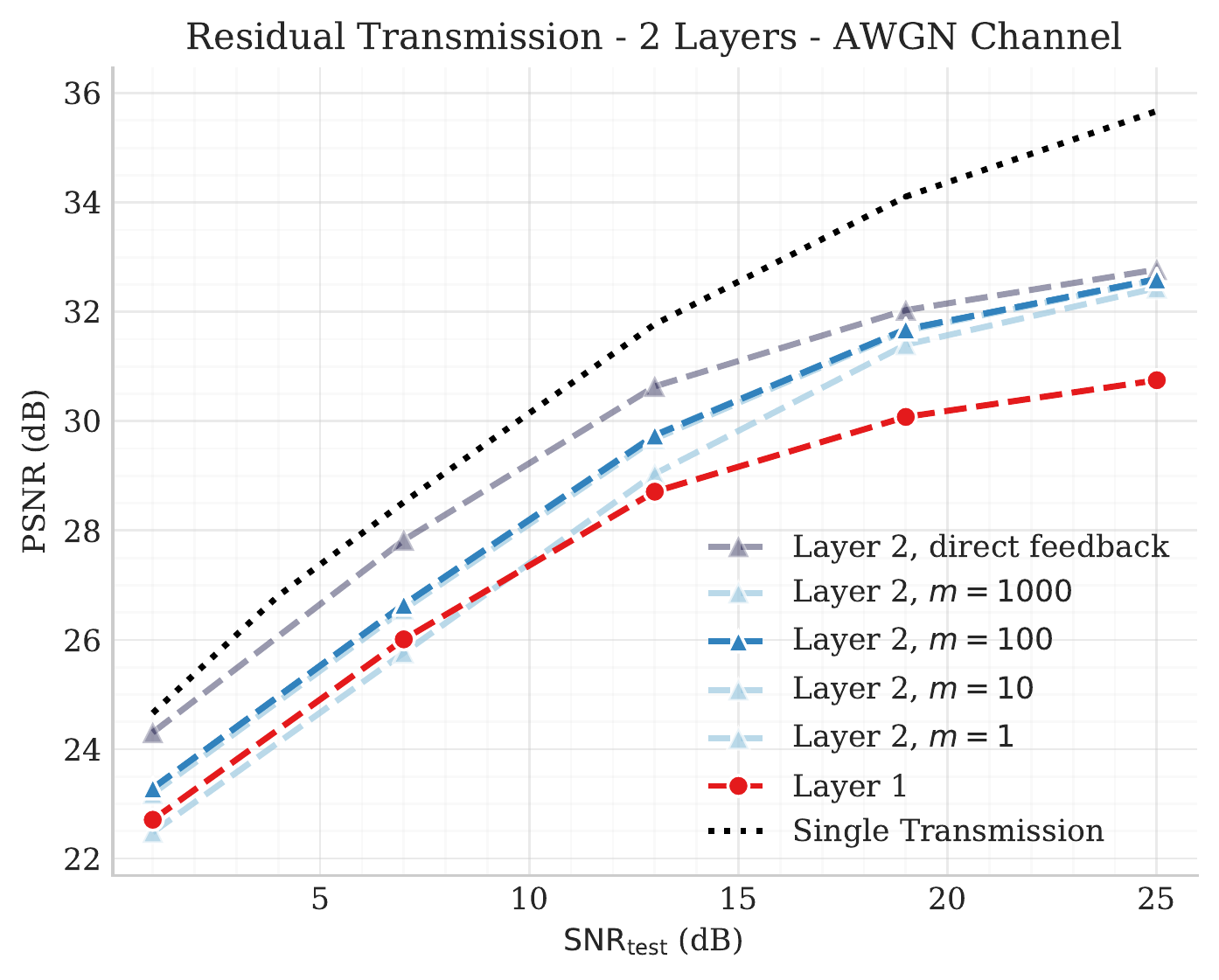}
 		\end{center}
	\vspace{-0.3cm}
		\caption{Performance of residual transmission scheme, AWGN channel, $k_1/n = k_2/n = 1/12$, with multiple values of $m$ considered.}\label{fig:resnet-awgn}
	\vspace{-0.3cm}
\end{figure}

Figure~\ref{fig:resnet-awgn} considers a similar scenario as
Figure~\ref{fig:dpass-awgn-05}, with AWGN channel and $k_1/n=k_2/n=1/12$. The results show that the scheme preserves the general advantages already presented (successive refinement, graceful degradation), but the performance is inferior.
One of the reasons that justify the lower performance is the fact that the residual being encoded is a function of the channel realization of the first transmission, which is unknown to the second encoder. Thus, the better the estimation of the decoded image of the previous layer is, the better is the quality of the residual transmission.

Figure~\ref{fig:resnet-awgn} presents the model's performance for different $m$ values, showing how the estimation accuracy increases with $m$, justifying the choice of $m=100$ as an appropriate value for practical implementations. As a reference, we also plot the performance of the (ideal) scenario in which the encoder has perfect channel output feedback; and hence, it can perfectly reconstruct the residual.
We see that the performance with perfect channel output feedback is closer to the one with two decoders trained jointly.
This is in line with the information theoretical results stating that feedback, in general, does not improve the end to end average reconstruction quality in this setting, but it can allow simpler more flexible schemes to be implemented.

\section{Single Decoder}

\begin{figure}[h]
	\begin{center}
        \resizebox {0.85\linewidth} {!}%
        {%

 \tikzstyle{txt} = [text centered]
 \tikzstyle{box} = [rectangle, rounded corners, minimum width=2.5cm, minimum height=1.5cm, text width=2.5cm, text centered, draw=black]
 \tikzstyle{bbox} = [rectangle, thick, minimum width=1.5cm, minimum height=1cm, text width=1.5cm, text centered, draw=black]
 \tikzstyle{arrow} = [thick,->,>=stealth]
 \tikzstyle{fitted} = [draw=gray, thick, dotted, inner sep=0.75em]

\begin{tikzpicture}[node distance=2.3cm]

\node (x) [txt, font=\fontsize{14}{0}\selectfont] {$\bm{x}$};
\node (encoder) [box, right of=x, font=\fontsize{12}{12}\selectfont] {Encoder \\ ($f_{\bm{\theta}}$)};
\node (z) [txt, right of=encoder, xshift=0.5cm, font=\fontsize{14}{0}\selectfont] { };
\node (channel1) [bbox, above right of=z,  font=\fontsize{12}{12}\selectfont, xshift=0.5cm, yshift=-0.5cm] {channel};
\node (decoder) [box, below right of=channel1, xshift=2.5cm, yshift=0.5cm, font=\fontsize{12}{12}\selectfont] {Decoder\\($g_{\bm{\phi}}$)};
\node (xhat1) [txt, above right of=decoder, xshift=1cm, font=\fontsize{14}{0}\selectfont] {$\bm{\hat{x}_1}$};

\node (channel2) [bbox, below right of=z,  font=\fontsize{12}{12}\selectfont, xshift=0.5cm, yshift=0.5cm] {channel};
\node (xhat2) [txt, below right of=decoder, xshift=1cm, font=\fontsize{14}{0}\selectfont] {$\bm{\hat{x}_2}$};

\draw [arrow] (x) -- (encoder);
\draw [arrow] (encoder.east)
-- node[above,font=\fontsize{14}{0}\selectfont] {$\bm{z}$} ++(1, 0) |- node[above right,font=\fontsize{14}{0}\selectfont] {$\bm{z_1}$}  (channel1.west);
\draw [arrow] (channel1) -- node[above,font=\fontsize{14}{0}\selectfont] {$\hat{\bm{z}}_1$} ++(2,0) |- (decoder);
\draw [arrow] (decoder) -| (xhat1);
\draw [arrow] (encoder.east)
-- ++(1, 0) |- node[below right,font=\fontsize{14}{0}\selectfont] {$\bm{z_2}$}  (channel2.west);
\draw [arrow] (channel2) -- node[below,font=\fontsize{14}{0}\selectfont] {$\hat{\bm{z}}_2$} ++(2,0) |- (decoder);
\draw [arrow] (decoder) -| (xhat2);

\end{tikzpicture}

}%
	\end{center}
\caption{Single decoder scheme with two layers.}
\label{fig:modelsingdec}
\end{figure}
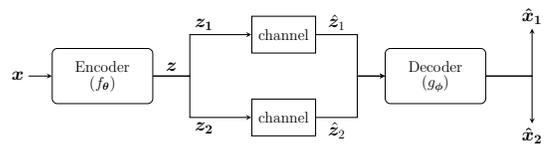

A downside of the model described in the previous sections is the fact that a separate encoder and/or decoder needs to be trained for each layer. Here we try an alternative model that uses a single encoder and a single decoder for all transmissions, as described in Figure~\ref{fig:modelsingdec}.
This represents a considerable reduction both in memory and in processing, as the model size is constant regardless of the number of layers.

In order to retrieve information from partial codes, the decoder has to be trained for different code sizes. We achieve that by
keeping a fixed loss function that compares inputs and outputs, while randomly varying the length of the code transmitted, at every batch.
In practical terms, that meant creating a CNN model with fixed channel bandwidth $k = \sum_{i=1}^L k_i$, but randomly masking regions of the received message $\hat{\bm z}$ with zeros. In this way, the network could learn to specialize different regions of the code, using the initial parts to encode the main image content and the extra (occasionally erased) parts for additional layers.

\begin{figure}[t]
	\begin{center}
 		\includegraphics[width=0.45\textwidth]{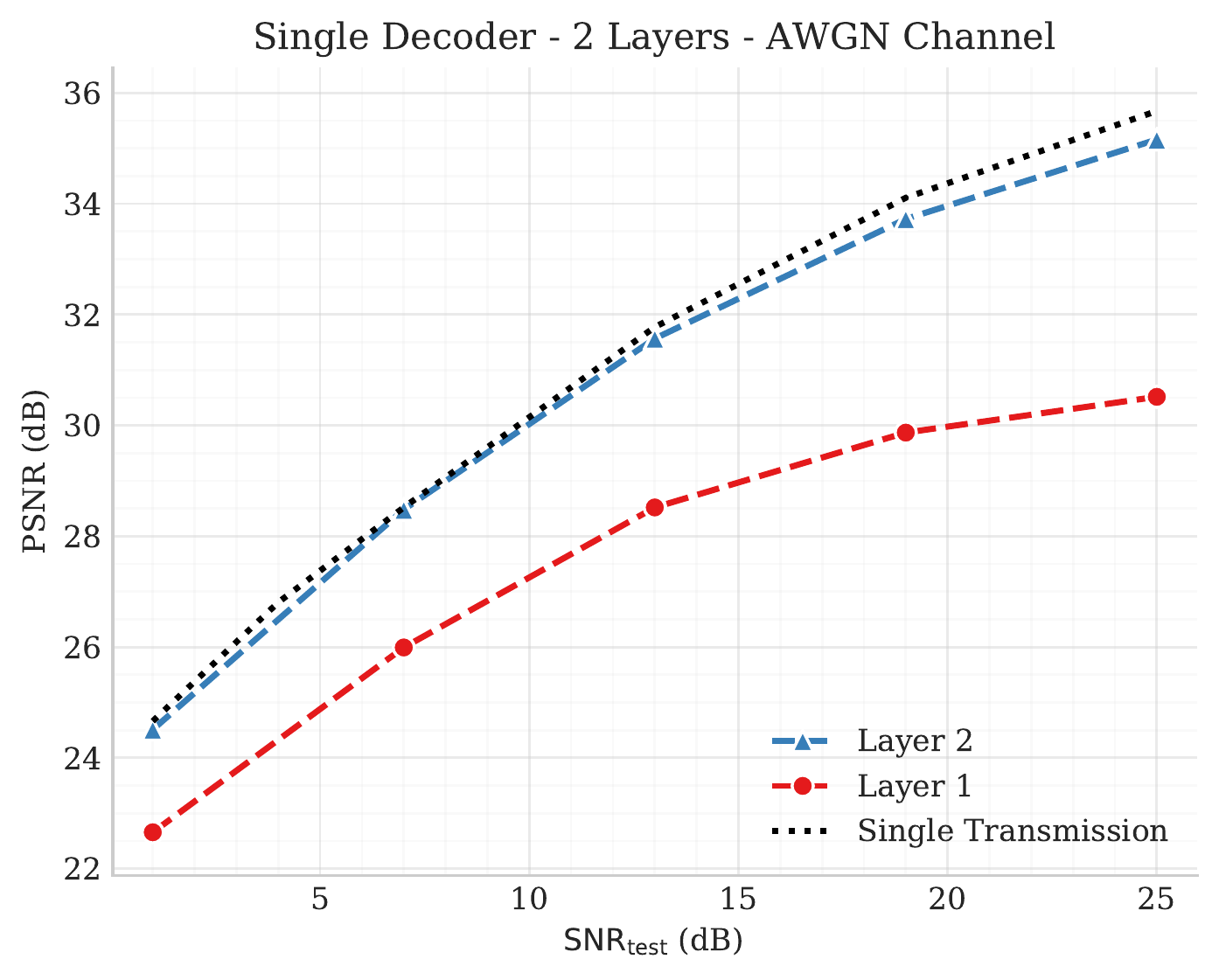}
 		\end{center}
	\vspace{-0.3cm}
		\caption{Performance of single decoder scheme, AWGN channel, $k_1/n = k_2/n = 1/12$.}
		\label{fig:single3}
		\vspace{-0.3cm}
\end{figure}

The results show that the performance of the single decoder scheme is remarkably powerful, as can be seen in the results presented in Figure~\ref{fig:single3}. The achieved values are comparable to the multiple decoder case, making this scheme appealing for models with small values of $L$.
The investigation of the impact of increasing $L$ in the scheme will be considered in further work.

\section{Conclusions}

We explored deep-learning based JSCC algorithms for successive refinement of images over wireless channels. To the best of our knowledge, no such hierarchical JSCC scheme has been previously developed and tested for practical information sources and channels.

We presented different strategies and models for progressive refinement - namely the use of multiple decoders, the transmission of residual images, and the use of a single encoder.
The results not only reproduce the effects observed in the previous work, such as impressive performance at low SNRs, limited bandwidth, and graceful degradation with test SNR, but also show the ability of neural networks in enabling progressive image transmission with almost no loss in the performance.

The best performance is obtained when a combination of one encoder and multiple receivers are trained jointly; however, alternative strategies such as the communication of residuals instead of complete images and the use of a single decoder also showed comparable results, being viable options depending on the deployed system's needs.

\section*{Acknowledgments}

This work was partially supported by the European Research Council (ERC) through project BEACON (No. 725731), by the European Union’s Horizon 2020 Research and Innovation Programme through project SCAVENGE (No. 675891) and by the National Council for Scientific and Technological Development (CNPq), Brazil.

\bibliographystyle{IEEEtran.bst}
\bibliography{layered.bib}

\end{document}